# EduBERT: Pretrained Deep Language Models for Learning Analytics

**Benjamin Clavié, Kobi Gal**
The University of Edinburgh
{Benjamin.clavie;kgal}@ed.ac.uk

**ABSTRACT**: The use of large pretrained neural networks to create contextualized word embeddings has drastically improved performance on several natural language processing (NLP) tasks. These computationally expensive models have begun to be applied to domain-specific NLP tasks such as re-hospitalization prediction from clinical notes. This paper demonstrates that using large pretrained models produces excellent results on common learning analytics tasks. Pre-training deep language models using student forum data from a wide array of online courses improves performance beyond the state of the art on three text classification tasks. We also show that a smaller, distilled version of our model produces the best results on two of the three tasks while limiting computational cost. We make both models available to the research community at large.[1]

## INTRODUCTION

In the past year, the field of Natural Language Processing (NLP) has seen the rise of pretrained language models such as as ELMo (Peters et al., 2018), ULMFiT (Howard and Ruder, 2018) and BERT (Devlin et al., 2019). These approaches train a deep-learning language model on large volumes of unlabeled text, which is subsequently fine-tuned for particular NLP tasks. Applying these models to the General Language Understanding Evaluation (GLUE) benchmark introduced by Wang et al. (2018) has achieved the best performance to date on tasks ranging from sentiment classification to question answering (Devlin et al., 2019).

The benefit of these models has also been demonstrated in specialized NLP domains. BioBERT (Lee et al., 2019), a version of BERT trained exclusively on biomedical text, was able to significantly increase performance on biomedical named entity recognition. Further refining this model on clinical text produced an increase in performance in medical natural language inference (Alsentzer et al. 2019).

While large pretrained models offer significantly increased performance, they come with their own constraints, as the number of parameters in the classic BERT-base model exceeds 100 million. As such, their computational cost can thus be prohibitively high at both training and prediction time (Devlin et al., 2019). More recent work has addressed this challenge by 'distilling' the models, training smaller versions of BERT which reduce the number of parameters to train by 40% while retaining more than 95% of the full model performance and even outperforming it on two out of eleven GLUE tasks (Sanh et al., 2019).

---

[1] Available at https://github.com/bclavie/edubert

This paper shows that using pretrained models in learning analytics holds great potential for advancing the field. We apply the BERT approach to the following three previously explored LAK tasks on MOOC forum data (Wei et al., 2017): Confusion detection, urgency of teacher intervention and sentimentality classification. In all three of these tasks, we are able to improve performance past the state of the art.

## METHOD

**Data:** We trained the language model on a large unannotated data set from two sources: student forum data from the Stanford MOOCPosts dataset (Agrawal and Paepcke, 2014) which includes about 30,000 forum posts from 11 courses among three subject domains; and forum data from multiple instances of 18 courses from large public universities in the UK and USA. In total, this dataset is comprised of more than 12 million tokens.

The data used for the classification tasks was from the same Stanford MOOCPosts dataset. The posts are annotated by domain experts and given scores for sentiment (the degree of emotionality exhibited by the post), confusion expressed by the student and urgency for the post to receive a response from an instructor. Scores are given on a Likert scale from 1 (low) to 7 (high).

**Language Models:** We constructed two models, EduBERT and EduDistilBERT, which respectively refine BERT-base and DistilBERT (Sanh et al., 2019), both of which were trained on general domain text from books and Wikipedia (Devlin et al., 2019). Both models are initialized from their base model and fine-tuned on educational data, using the Transformers library (Wolf et al., 2019). The fine-tuning step allows the model to better capture how words are used in an educational context.

Training of the models was performed on a Titan X GPU. We set the maximum input sequence length to the default value (512); the learning rate was set to 5e-5; the batch size (the number of input sequences processed at one time) was set to 8 for EduBERT and 16 for EduDistilBERT. The best performance was achieved after 5 training epochs.

**Classification Tasks:** To encourage easily comparable results, we evaluated the models on three well-explored classification tasks on the StandfordMOOC dataset. Following previous work by Guo et al. (2019), we split the data into a 2/3 training set and 1/3 test set and consider a post to express sentiment, urgency or confusion if and only if its respective score is ≥ 4.

We compare between the four classifiers BERT-base, DistilBERT, EduBERT and EduDistilBERT. We evaluated multiple sets of parameters. Best results for these tasks were achieved with the following parameters: two learning epochs, maximal sequence length of 300 (BERT-base, EDUBERT) and 512 for the distilled models, all other parameter values were equal to the ones used for pre-training.

## RESULTS & DISCUSSIONS

**Table 1** compares EduBERT, EduDistilBERT to their base versions, as well as the state-of-the-art (SoA) for urgency detection (Guo et al. 2019). The table shows that all pretraining approaches outperformed the SoA for F1 and weighted F1 measures, with our distilled model EduDistilBERT achieving the best overall performance. **Table 2** compares all of the models for all three tasks to the SoA using the same measures of accuracy as Wei et al. (2017). Here too, all the pretraining approaches outperform the SoA. EduDistilBERT obtains the best results on both urgency and confusion prediction while EduBERT performs the best for sentimentality classification. However, EduDistilBERT has a lower memory footprint and is noticeably faster at inference time, allowing for a 30% speedup.

.

Table 1: Performance metrics for urgency prediction

|  | Non-urgent | | | Urgent | | | |
| --- | --- | --- | --- | --- | --- | --- | --- |
|  | Recall | Precision | F1 | Recall | Precision | F1 | Weighted F1 |
| EduDistilBERT (Ours) | 0.949 | **0.954** | **0.952** | 0.835 | 0.819 | **0.827** | **0.925** |
| DistilBERT | 0.946 | 0.953 | 0.950 | 0.833 | 0.810 | 0.821 | 0.921 |
| EduBERT (Ours) | 0.950 | 0.950 | 0.950 | 0.822 | 0.820 | 0.821 | 0.922 |
| BERT-base | **0.956** | 0.944 | 0.950 | 0.794 | **0.835** | 0.814 | 0.920 |
| Guo et al. (2019) | 0.954 | 0.948 | 0.951 | 0.772 | 0.834 | 0.801 | 0.918 |

Table 2: Accuracy measures on the three tasks

|  | Confusion | Sentiment | Urgency |
| --- | --- | --- | --- |
| EduDistilBERT | **83.01** | 89.67 | **92.43** |
| DistilBERT | 82.88 | 89.12 | 92.14 |
| EduBERT | 82.91 | **89.78** | 92.24 |
| BERT-base | 82.80 | 89.47 | 92.14 |
| Wei et al. (2017) | 81.88 | 86.08 | 86.68 |

**Future Work & Conclusion.** EduBERT and EduDistilBERT are fine-tuned on millions of tokens, in contrast to the billions of tokens required to make the most of the architecture potential (Devlin et al., 2019). We are actively seeking more data to train models even more capable of producing contextualized word representations in the educational domain. We are making EduBERT and EduDistilBERT publicly available in the hope that they will facilitate learning analytics research at large.